\documentclass[a4paper]{IEEEtran}
\ifCLASSINFOpdf
  \usepackage[pdftex]{graphicx}
\else
\fi

\begin{document}
%
\title{Predicting proximity with ambient mobile sensors for non-invasive health diagnostics}

\author{\IEEEauthorblockN{Sylvester Olubolu Orimaye, Foo Chuan Leong, Chen Hui Lee, Eddy Cheng Han Ng}\\
\IEEEauthorblockA{Intelligent Health Research Group,  School of Information Technology\\
Monash University Malaysia\\
Email: sylvester.orimaye@monash.edu}
}


%


\maketitle

\begin{abstract}
Modern smart phones are becoming helpful in the areas of Internet-Of-Things (IoT) and ambient health intelligence. By learning data from several mobile sensors, we detect nearness of the human body to a mobile device in a three-dimensional space with no physical contact with the device for non-invasive health diagnostics. We show that the human body generates wave patterns that interact with other naturally occurring ambient signals that could be measured by mobile sensors, such as, temperature, humidity, magnetic field, acceleration, gravity, and light. This interaction consequentially alters the patterns of the naturally occurring signals, and thus, exhibits characteristics that could be learned to predict the nearness of the human body to a mobile device, hence provide diagnostic information for medical practitioners. Our prediction technique achieved 88.75\% accuracy and 88.3\% specificity.
\end{abstract}


%
\IEEEpeerreviewmaketitle

\section{Introduction}
The importance of predicting with mobile sensors can be seen in various mobile apps that help people's lifestyle. In most cases, the \emph{acceleration} sensor in mobile devices have been used to capture activity patterns using the x, y and z coordinate values generated by the sensor. This has given substantial accuracy, especially for activities that generate vivid alternating patterns, such as running and jogging \cite{kwapisz:2011}. However, it is rather unclear whether other sensors could as well contribute to the prediction, and perhaps, the combination of two or more sensors could lead to better prediction of different activities.

In this paper, we predict a novel activity that has to do with the nearness or proximity distance of a human body to the mobile device without physically touching or holding the mobile device. As such, for the purpose of this work, we refer to proximity distance recognition as \emph{``nearness recognition"}, and hence, will be used consistently for the rest of this paper.

Human-mobile nearness recognition is important and could significantly change the way humans communicate with their mobile devices. Nearness recognition could be used in the area of \emph{ambient health intelligence}, whereby human health information such as wellness or exposure to environmental hazards can be autonomously detected by a mobile app. For example, \cite{jafari:2005} used the ambient temperature and pressure sensors to measure a user's physiological conditions and disorders in hazardous environments. High humidity can cause mold and fungus to grow and affect people with asthma and allergies, while in low humidity, dry skin and eyes itchiness could develop \cite{hoffmann:2012}. Thus, detecting these ambient changes around humans could potentially prevent health problems.

\begin{figure}[h]
\begin{center}
\includegraphics[width=75mm,height=40mm]{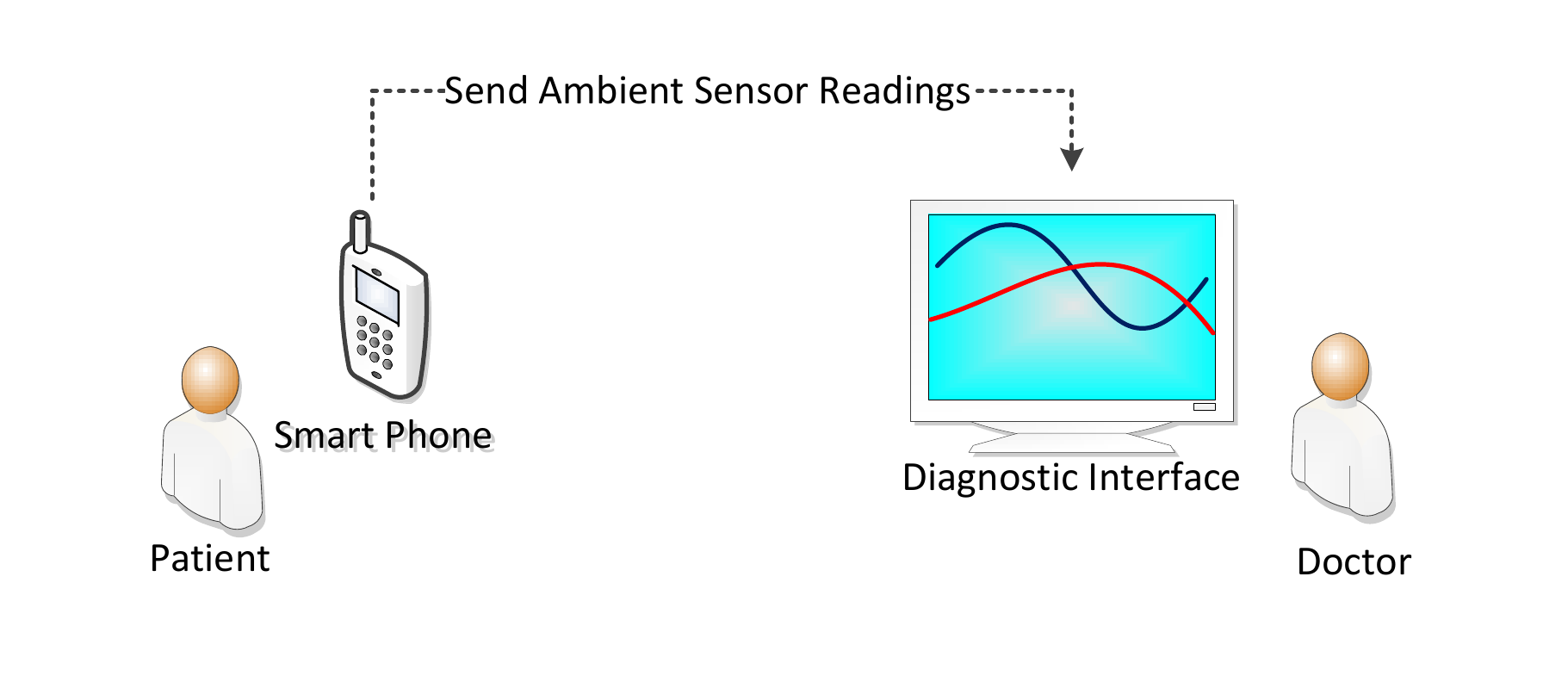}
\caption{\label{fig:overview} Using mobile ambient sensors for health diagnostics.}
\end{center}
\end{figure}

As such, we perform experiments by collecting data from \emph{steady} and \emph{non-steady} spaces and perform several data transformations to show that both exhibit different patterns or signatures, and thus can be studied and used as features to train machine learning algorithms. Second, we predict that a human is near to a mobile device by up to 88.75\% accuracy (88.3\% sensitivity) and then collect several ambient information for diagnostic purposes as shown in Figure \ref{fig:overview}.

The rest of this paper is organized as follows. Related works are discussed in Section \ref{section:RelatedWork}; \emph{steady} and \emph{non-steady} spaces in Section \ref{section:SteadyNonSteady}; nearness recognition in Section \ref{section:NeanessRecognition}; experiments and results in Section \ref{section:Experiments}; and conclusion in Section \ref{section:ConclusionFutureWork}.

\section{Related Work}
\label{section:RelatedWork}
Most related works have performed activity recognition for healthy lifestyle and fitness using the acceleration sensor \cite{kwapisz:2011,brezmes:2009}. For example,\cite{kwapisz:2011} performed activity recognition using the acceleration sensor on Android based mobile device. Several activities including walking, jogging, standing, sitting, and ascending or descending stairs were predicted having learned a set of transformed features. Acceleration data was transformed into 43 learn-able features. Multilayer perceptron has the best predictive accuracy of 91.7\% followed by J48 with 85.1\%. Our work differs from \cite{kwapisz:2011} as we predict nearness or noticeable proximity of the human body to a mobile device. 

More recently, proximity detection has been introduced for creating several non-invasive health diagnostic tool for the health industry using Body Area Network (BAN) techniques \cite{chen:2011}. Our work falls in this category, and more importantly, novel in the sense that we do not use only one sensor point or multiple devices in our prediction. We combined multiple sensors' data on one mobile device alone to improve the proximity prediction accuracy.

\section{Steady and Non-Steady Spaces}
\label{section:SteadyNonSteady}

\begin{figure}[h]
\begin{center}
\includegraphics[width=45mm,height=15mm]{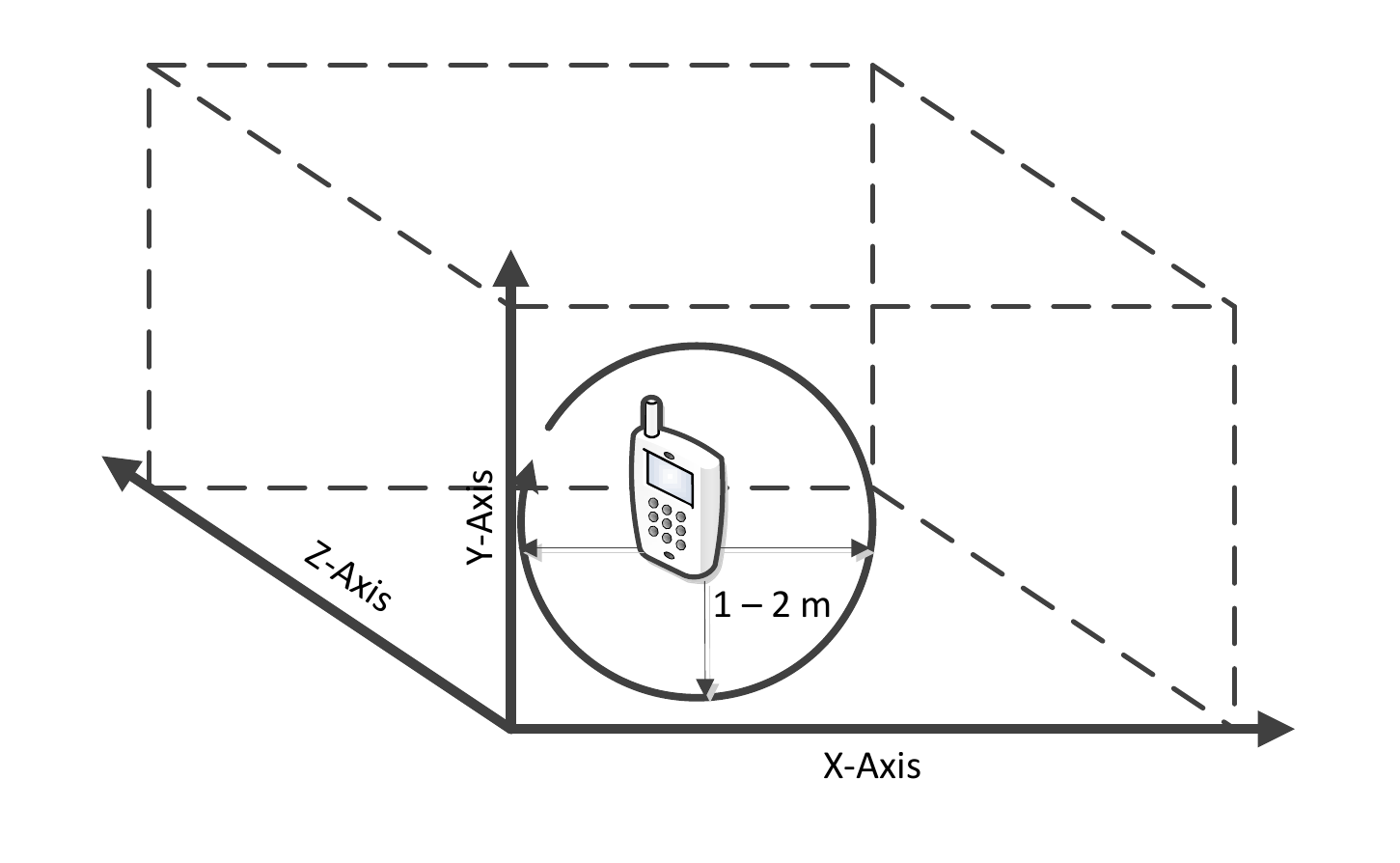}
\caption{\label{fig:steady} A representation of a steady three-dimensional space having no human body interaction at a time interval $t_{1}-t_{n-1}$.}
\end{center}
\end{figure}

\begin{figure}[h]
\begin{center}
\includegraphics[width=45mm,height=15mm]{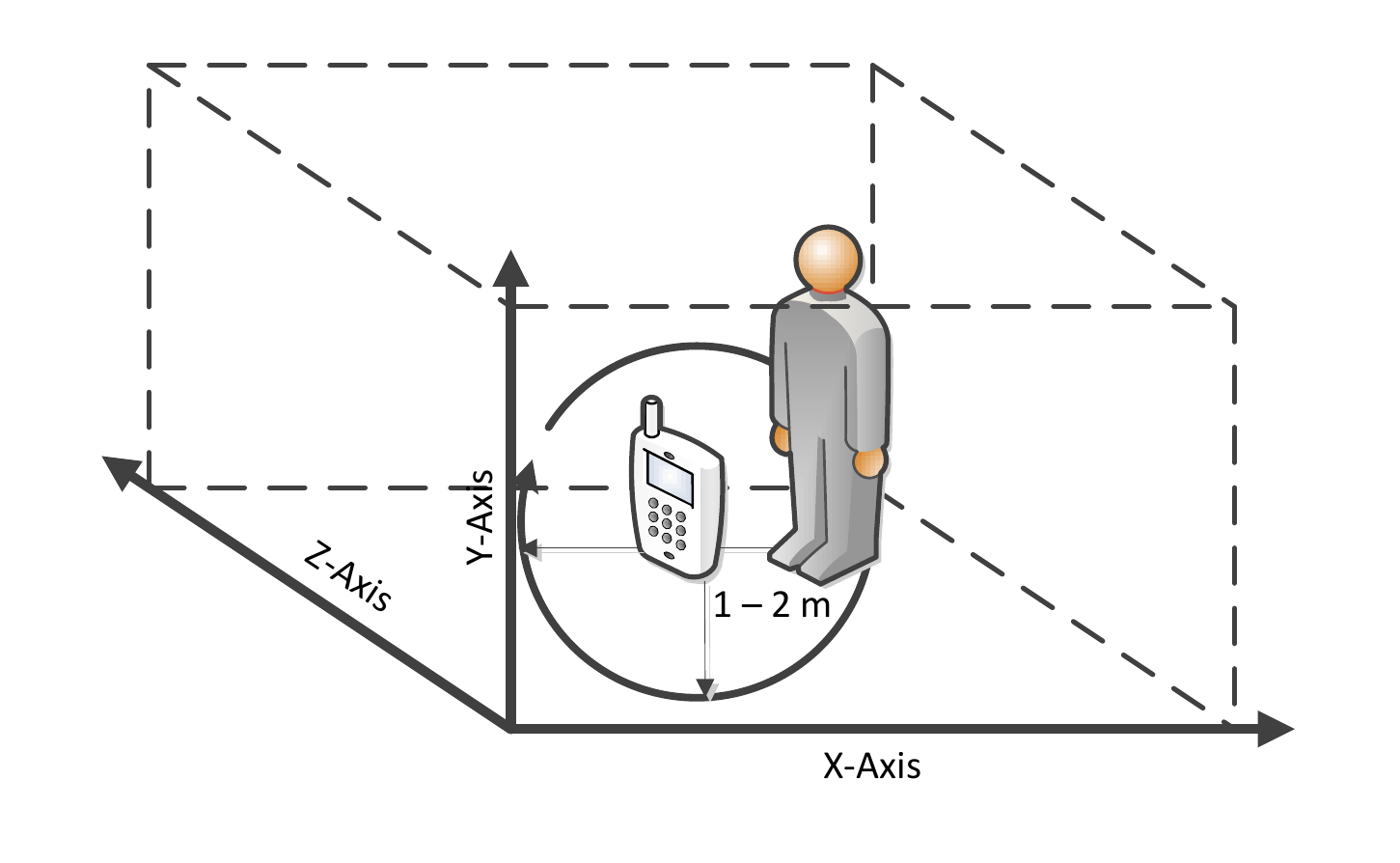}
\caption{\label{fig:nonsteady} A representation of a non-steady three-dimensional space with human body interaction at a time interval $t_{1}-t_{m-1}$.}
\end{center}
\end{figure}

In this work, we identify a three-dimensional space as a habitable area that supports physical interactions between a human body and other objects therein. Steady and non-steady spaces can be likened to the popular \emph{steady and transient states} in thermodynamics \cite{okoro:2005}, and audio signal processing \cite{mu:2012}, whereby the output of a filter generates unsteady sinewave before resulting to a steady sinewave of similar frequency, see Figures \ref{fig:steady} and \ref{fig:nonsteady}. 

Similarly, we propose that, at a certain \emph{spatio-temporal} window $S_{w}$, a space could be in a steady state within a particular time interval $t_{1}-t_{n-1}$, and subsequently enters into a non-steady space in another time interval $t_{1}-t_{m-1}$ due to a sudden interference caused by a human body on the steady space.

\subsection{Steady Space}
\begin{figure}[h]
\begin{center}
\includegraphics[width=70mm,height=40mm]{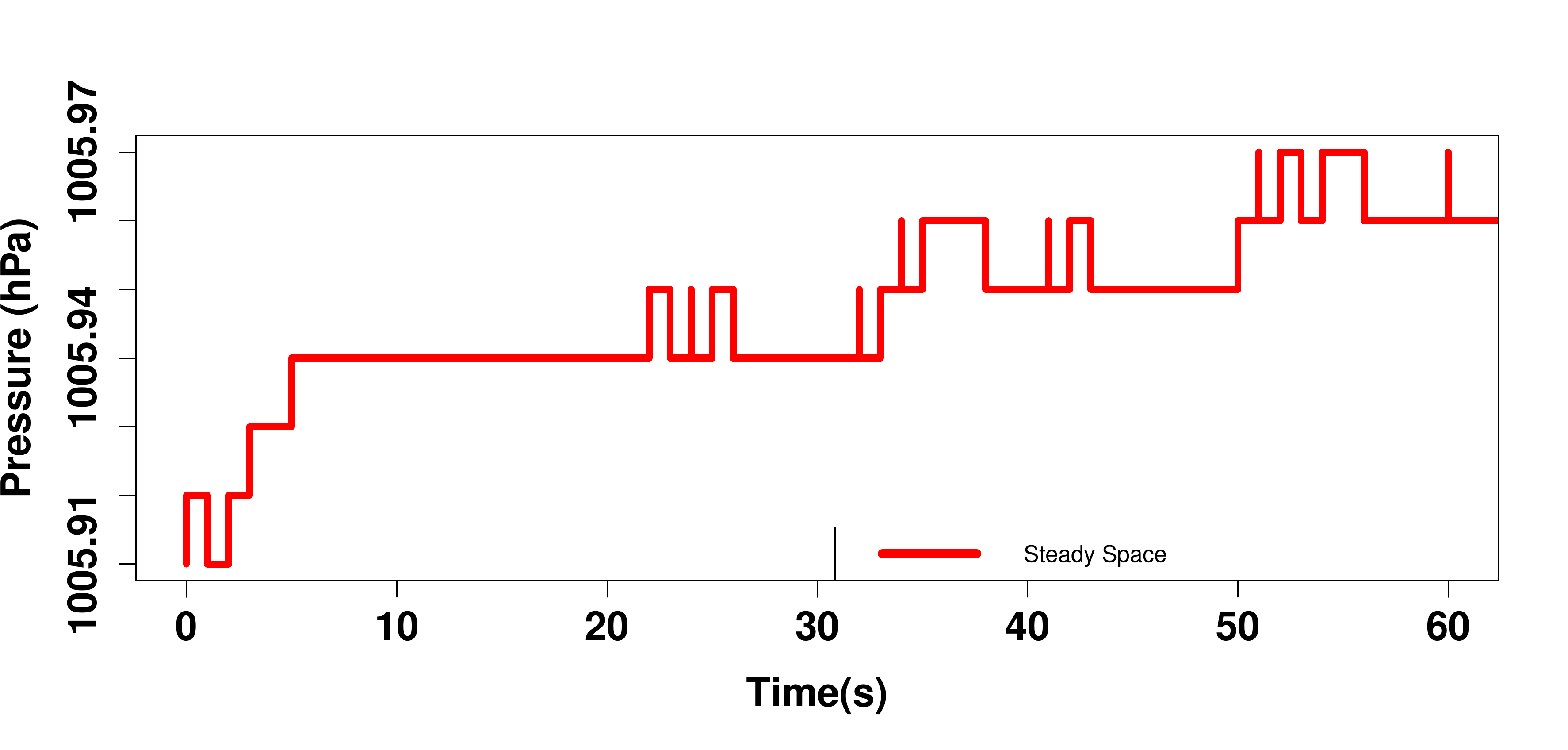}
\caption{\label{fig:steadySinusoid} An example of a steady three-dimensional space sinusoidal output for a pressure sensor with T=60s.}
\end{center}
\end{figure}

We refer to a steady space as a three-dimensional space around a mobile device with measurable sinusoids at a specific period $T$. The steady state excludes the presence or existence of a human body at a close proximity to the mobile device as illustrated in Figure \ref{fig:steady}. Figure \ref{fig:steadySinusoid} shows a steady three-dimensional space with the sinusoidal output of a pressure sensor within a 60 seconds period (i.e. $T = 60s$). 

As such, we define the characteristic of a steady space as 

\begin{equation}
f(t) = \sum_{i=1}^{S_{n}} a\sin \omega t + \lambda 
\end{equation}

where $S_{n}$ is the number of sinusoidal signals present, $a$ is the \texttt{constant} amplitude of each sinusoid, $\omega$ is the frequency of each sinusoid, and $\lambda$ is a normalization parameter for each steady space signal.

\subsection{Non-Steady Space}

\begin{figure}[h]
\begin{center}
\includegraphics[width=70mm,height=40mm]{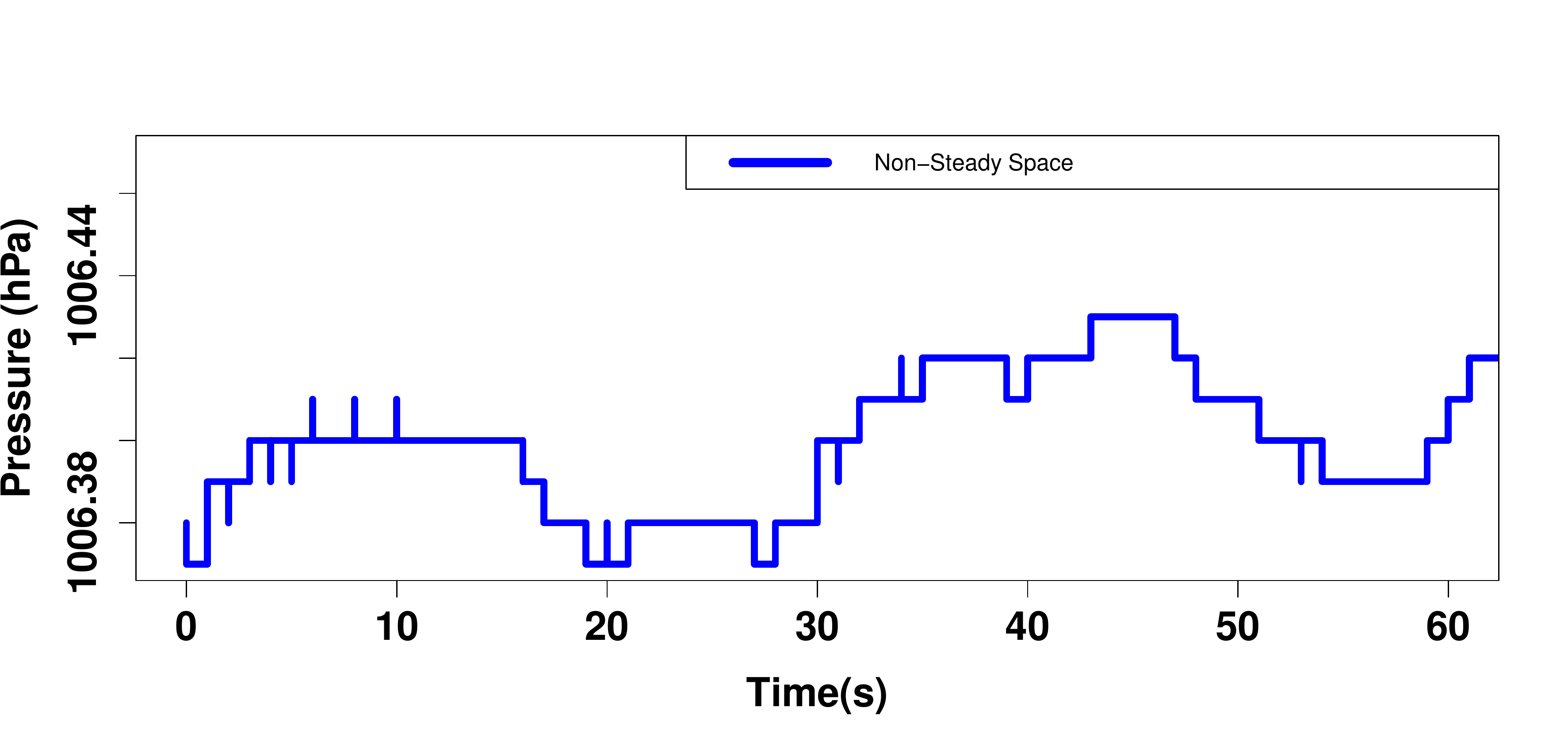}
\caption{\label{fig:nonsteadySinusoid} An example of non-steady three-dimensional space sinusoidal output for a pressure sensor with $T= 60s$.}
\end{center}
\end{figure}

We identify a non-steady space as a three-dimensional steady space that enters into \emph{transient} as a result of the extra oscillation and radiation introduced to the steady space with an entry of a human body. Thus, Figure \ref{fig:nonsteadySinusoid} shows a non-steady three-dimensional space, again with the sinusoidal output of a pressure sensor within a 60 seconds period (i.e. $T = 60s$).

Thus, we define the characteristic of a non-steady space as

\begin{equation}
\label{eqn:nonsteady}
f(t) = \sum_{i=1}^{S_{n} + H} \gamma(t) \sin \omega t + \lambda 
\end{equation}

where $H$ is the extra human oscillation or signal introduced to the set of existing sinusoidal signals,$\gamma(t)$ is the transient component introduced as a result of the introduction of $H$, $\omega$ is the frequency of each sinusoid, and $\lambda$ is a normalization factor for the sinusoids.

\section{Nearness Recognition}
\label{section:NeanessRecognition}
Having defined the characteristics of steady and non-steady spaces, we propose to learn the characteristics in order to detect human-mobile nearness. We propose that a human is likely to be near or in a close proximity to a mobile device when the condition in Equation \ref{eqn:nonsteady} is satisfied. Moreover, signals that are characterized by Equation \ref{eqn:nonsteady}, have also been characterized by a similar and popular equation in speech signal processing \cite{swami:1994}:

\begin{equation}
\label{eqn:random}
F(t) = g(t) e^{j \omega t} + h(t), t= 0,...n-1 
\end{equation}

where $g(t)$ is the random amplitude, $h(t)$ is a noise component, and $\omega$ is defined as the transient frequency. In order to detect human-mobile nearness, one could learn the $g(t)$, $\omega$, and $h(t)$ components \cite{mu:2013}. In addition, research have shown that the human body is capable of generating waveforms as a result of the endogenous electromagnetic fields which radiate from the human body to the three-dimensional space \cite{haron:2012,cho:2007}. In a recent research by \cite{mackowiak:2011}, it was shown that the human body could significantly alter the electromagnetic waves that are generated by sensors, thereby leading to noisy wave pattern. 

\subsection{Data Collection}
\label{section:DataCollection} 

We developed an Android application to log several sensor readings from a Samsung Galaxy S4 mobile device in a controlled environment. The sensors include, Pressure, Gravity, Accelerometer, Ambient Temperature, Relative Humidity, Magnetometer, Gyroscope, Ambient Light, and Rotation Vector. The data logged per one reading instance include the X,Y,Z coordinates of each sensor and the reading timestamp in milliseconds.

Our controlled environment is a designated behavioral lab, which was designed for the purpose of conducting behavioral experiments relating to humans. Thus, we regard the behavioral lab as a three-dimensional space as discussed in Section \ref{section:SteadyNonSteady}. Since we propose to learn both steady and non-steady spaces, we performed data collection in two phases. 

In phase 1, a three-dimensional \emph{steady} space is simulated by starting the data collection app with the mobile device placed on a table at the common area within the lab. In this phase, no human presence is required in the lab. We then log readings from the 9 sensors over a period of 60 minutes. 

In phase 2, we simulated a three-dimensional \emph{non-steady} space by maintaining the setup in phase 1 and include a human subject to walk or sit around the mobile device up to a distance of 1 meter from it over a period of 10 minutes while logging the sensor readings.

Phase 2 was performed with 20 different participants who are regular mobile phone users. Furthermore, the total reading instances per phase from each human subject was divided into several \emph{reading windows}. Each 10 minute period in a phase, generated approximately 6000 reading instances on average. We limit each reading window to a maximum of 512 in order to allow the sinusoidal outputs generate enough unique and learn-able characteristics as suggested in \cite{mu:2013}.  

\subsection{Data Transformation}
\label{section:DataTransformation}
It is essential to transform the raw time-series readings to generate different learn-able features \cite{mu:2013}. As such, we use several statistical and numerical techniques to transform the data for each sensor. 

\begin{itemize}
\item \textbf{Mean ($\mu$) \cite{kwapisz:2011}:} average of each coordinate X,Y, and Z over the total number of reading instances in a reading window.
\begin{equation}
\label{eqn:mean}
\mu(X,Y,Z) = \frac{\sum_{i=1}^{N} r_{i}}{W_{r}}  
\end{equation}
where $N$ is the number of readings in each coordinate, $r$ is a reading value, and $W_{r}$ is the reading window set to 512.
\item \textbf{Standard Deviation (SD) \cite{kwapisz:2011}:} the standard deviation for each coordinate X,Y, and Z over the total number of reading instances in a reading window. 
\begin{equation}
\label{eqn:SD}
SD(X,Y,Z) = \sqrt{\frac{1}{N}\sum_{i=1}^{N}(r_{i}-\mu)^{2}}   
\end{equation}

where $N$ is the number of readings in each coordinate, $r$ is a reading value, and $\mu$ is the mean of the reading values.
\item \textbf{Entropy($\eta$):} the entropy for each coordinate X,Y, and Z over the total number of reading instances in a reading window. Entropy measures the \emph{level of uncertainty} in the sample data, where the entropy is minimal for a less random data and larger for a more random data \cite{gray:2011}.

\begin{equation}
\label{eqn:entropy}
\eta(X,Y,Z) = \frac{1}{2} \ln(2 \pi e \sigma^{2} ) 
\end{equation}

where X,Y,Z are sets of reading instances for the coordinates, $e$ is an exponential function, and $\sigma^{2}$ is the \emph{variance}.

\item \textbf{Mean Absolute Difference (MAD) \cite{kwapisz:2011}:} For each X,Y,Z coordinates in a sensor, we compute the mean absolute difference as the mean of the difference between the values in each coordinate and the mean value for that coordinate.

\begin{equation}
\label{eqn:MAD}
MAD(X,Y,Z) = \frac{\sum_{i=1}^{N} r_{i}-\mu}{W_{r}}
\end{equation}

\item \textbf{Mean Resultant Weight(MRW) \cite{kwapisz:2011}:} The weight is computed as the square root of the sum of the square of each value in the X,Y,Z coordinates divided by the number of reading instances in the reading window.
\begin{equation}
\label{eqn:MRW}
 MRW_{XYZ} =  \sqrt{\frac{\sum_{i=1}^{N}x_{i}^{2}+y_{i}^{2}+z_{i}^{2}}{W_{r}}} 
\end{equation}
where $x, y$ and $z$ are reading values from X,Y,Z coordinates, and $W_{r}$ is the reading window.

\item \textbf{Gaussian Coverage Strength (GCS):}
Gaussian distribution has been successfully used in signal processing for interpreting random variables \cite{davenport:2010}. Similarly, we transformed the values of the X, Y, Z coordinates with the Gaussian's Probability Density Function (PDF) \cite{parzen:1962}, and the Cumulative Distribution Function (CDF) \cite{smirnov:1948}, to represent the Gaussian distributions, respectively. The PDF is computed as follows:

\begin{equation}
\label{eqn:PDF}
 PDF =  \frac{1}{\sigma \sqrt{2 \pi}}e^{- \frac{(r_{i}-\mu)^2}{2 \sigma ^ 2}} 
\end{equation}

where $\sigma$ is the standard deviation, $e$ is an exponential function, $r_{i}$ is each reading, $\mu$ is the mean, and $\sigma^2$ is the variance. Similarly, we compute the CDF as follows:

\begin{equation}
\label{eqn:CDF}
 CDF = \frac{1}{2}[1 + erf (\frac{r_{i} - \mu}{\sigma \sqrt{2}}) ]
\end{equation}

where $erf$ is the Gaussian error function for each reading $r_{i}$ \cite{chang:2011}, thus computed as:

\begin{equation}
\label{eqn:erf}
 erf(r) = \frac{2}{\sqrt{\pi}} \int_{0}^{r}  e^{-t^{2}} dt
\end{equation} 

Figures \ref{fig:mag-CDF} shows the differences in the Gaussian transformations (CDF) of the X-axis of a magnetometer sensor readings.

\begin{figure}[h]
\begin{center}
\includegraphics[width=70mm,height=40mm]{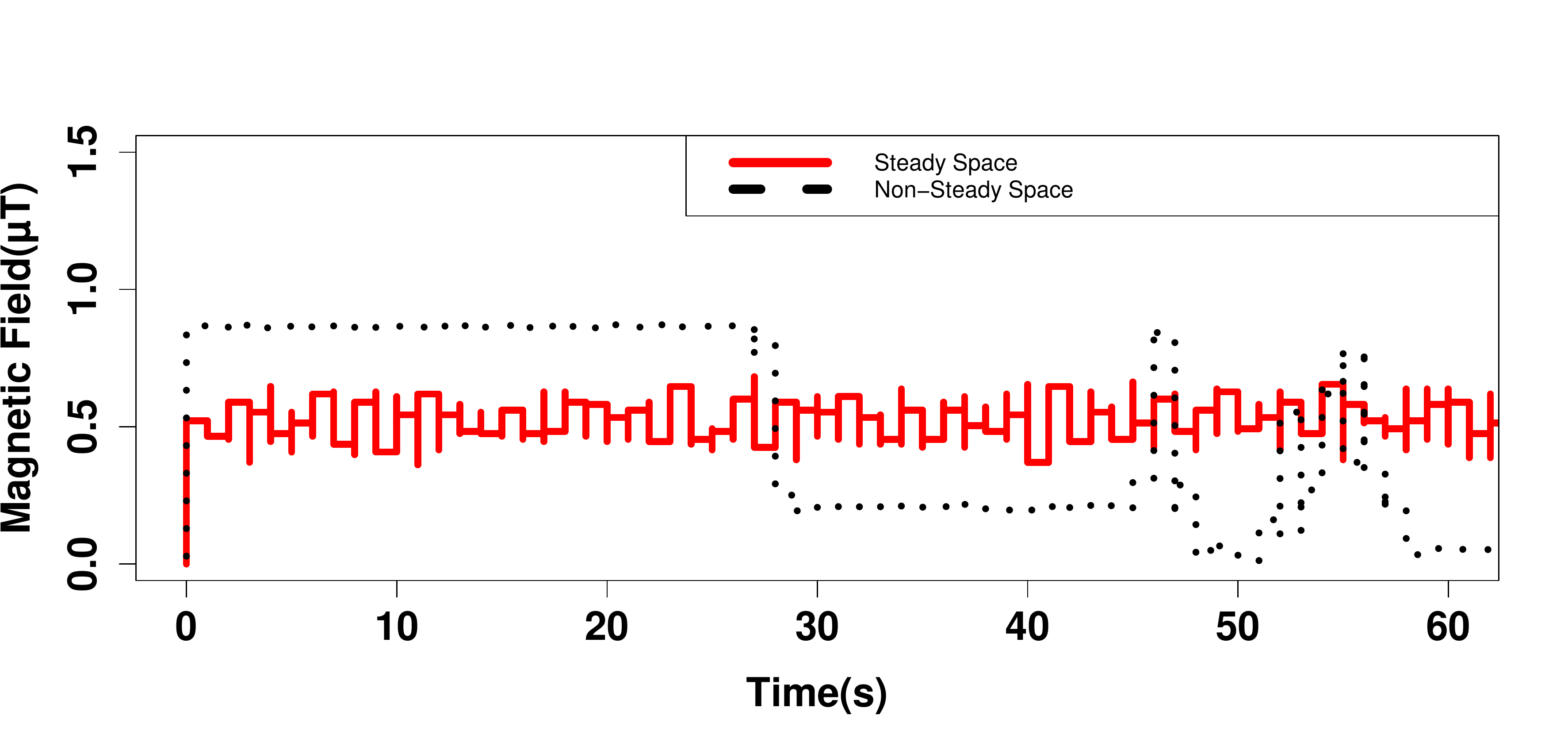}
\caption{\label{fig:mag-CDF} Difference between the CDF transformation for the X-axis of a magnetometer sensor readings from steady and non-steady spaces.}
\end{center}
\end{figure} 

Hence, the GCS is computed as the difference between the maximum and minimum values of each of the Gaussian distributions for X, Y, Z coordinates as follows:

\begin{equation}
\label{eqn:GCS}
 GCS = \max (D_{Gaus}) - \min (D_{Gaus})
\end{equation}
where $D_{Gaus}$ is the Gaussian distribution in PDF or CDF.

\item \textbf{Gaussian Average Peak Intervals (GAPI):} the average peak-to-peak intervals of each sinusoidal output generated by the Gaussian distributions. The interval is computed as the difference between the log of the first peak $pk_i$ to the next peak $pk_{i+1}$. GAPI uses the CDF and PDF functions for the the X,Y,Z coordinates as follows:

\begin{equation}
\label{eqn:GAPI}
 GAPI = \frac{\sum_{i=1}^{N} \log(pk_{i}) - \log (pk_{i+1})}{W_r} 
\end{equation}

where $pk$ is a peak in the sinusoid and $W_r$ is the reading window as discussed earlier.

\item \textbf{Fast Fourier Transform Average Distinct Peak Interval(FFT-ADPI):}
As in \cite{mu:2013}, we use the FFT to transform the values from the X,Y,Z coordinates of each sensor readings and then characterize the sinusoidal output using the resulting FFT magnitudes. We then compute the mean of the magnitudes for each axis and then find the average of the intervals between the peaks that are greater than or equal to the mean of the magnitudes. Peaks that are greater than or equal to the mean are regarded as \emph{Distinct Peaks}.

\begin{equation}
\label{eqn:FFT-ADPI}
 ADPI = \frac{\sum_{i=1}^{N} pk_{i} - pk_{i+1}}{W_r}; \\   pk_{i} \geq \mu (mag_{FFT})
\end{equation}

where $\mu (mag_{FFT})$ is the mean of the magnitudes generated by the FFT. Note that we did not compute the \emph{interval} between the distinct peaks as the difference between the ``log" of the peaks as in Equation \ref{eqn:GAPI}. This is because the FFT generates relatively smooth graphs or patterns (see Figure \ref{fig:fft-CDF}), which makes the magnitudes of its sinusoids very small \cite{mu:2013}, and thus, generates peaks that are very close to the origin. As such, the log of such peaks could lead to ambiguous intervals. Since the CDF generates visible regular patterns with more distinct peaks (see figures \ref{fig:mag-CDF}), our FFT-ADPI is computed on the CDF function over X,Y,Z axis of each sensor. 

\begin{figure}[h]
\begin{center}
\includegraphics[width=70mm,height=40mm]{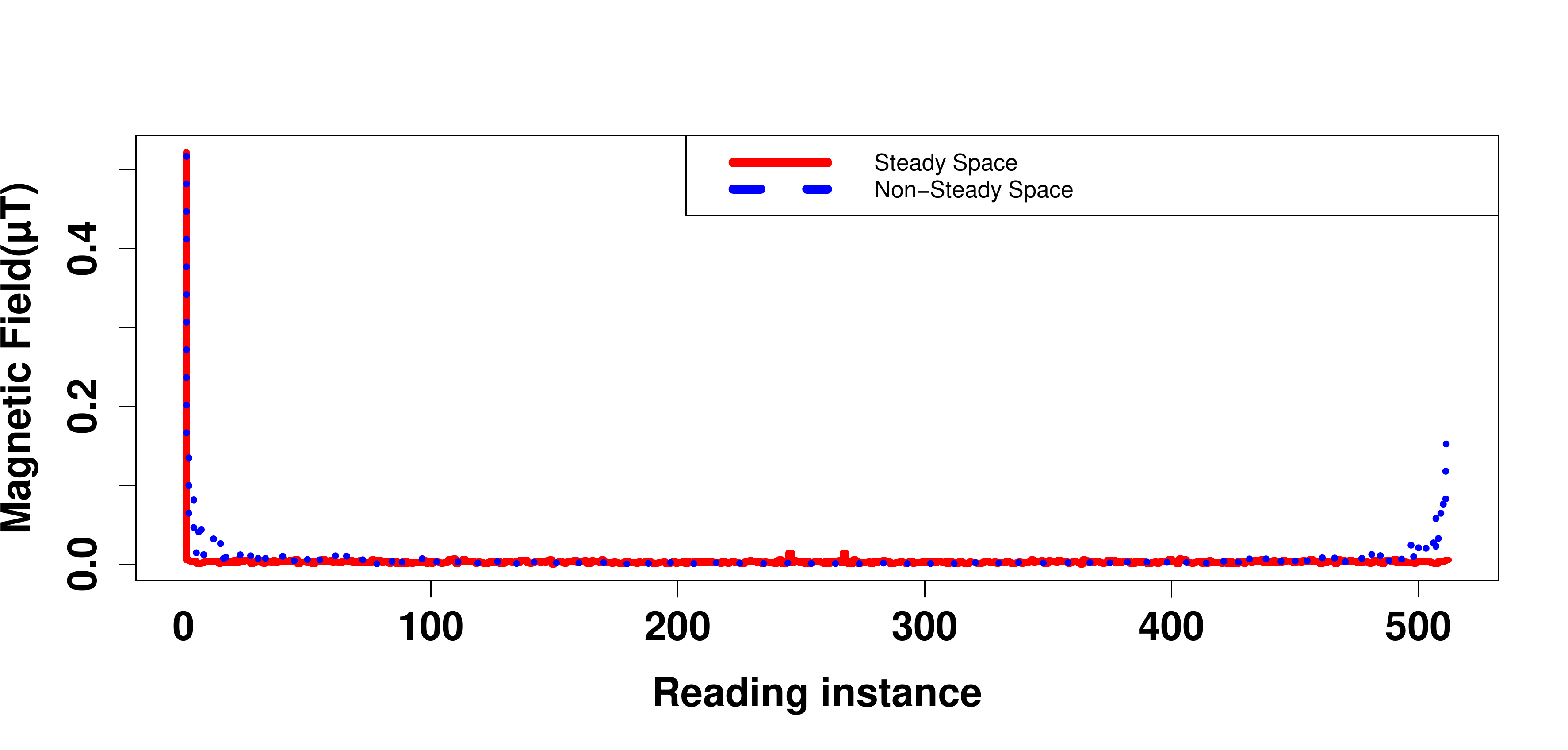}
\caption{\label{fig:fft-CDF} FFT-CDF transformation for the X-axis of a magnetometer sensor readings from steady and non-steady spaces over a 512 reading window.}
\end{center}
\end{figure} 

\end{itemize} 

\subsection{Features Set}
Table \ref{table:features-set} shows the combined features set resulting from the transformation processes. The set will be used in training different machine learning algorithms for predicting the nearness of a human to a mobile device.

\begin{table*}[ht]
\begin{center}
\caption{\label{table:features-set}Combined features set from each sensor for learning ML algorithms}
\begin{tabular}{p{0.5cm} p{1.5cm} p{5.3cm} p{3.5cm} } 
\hline
\textbf{SN}&\textbf{Feature}&\textbf{Description/Summary}&\textbf{Symbols}\\ 
\hline
1 & Mean & Average over the values in each X,Y,Z coordinate. & Mean-X, Mean-Y, Mean-Z.\\ 

2 & SD & Standard Deviation over the values in each X,Y,Z coordinate. & SD-X, SD-Y, SD-Z.\\ 
3 & Entropy & Level of uncertainty over the values in each X,Y,Z coordinate. & Entropy-X, Entropy-Y, Entropy-Z.\\ 
4 & MAD & Mean of the difference between each X,Y,Z coordinate value and their means. & MAD-X, MAD-Y, MAD-Z.\\ 
5 & MRW & Square of the sum of square roots of X,Y,Z coordinate values divided by reading window. & MRW.\\
6 & GCS-CDF & Cumulative Distribution Function over each X,Y,Z coordinate values. & GCS-CDF-X, GCS-CDF-Y, GCS-CDF-Z.\\
7 & GAPI-PDF & Average of peak-to-peak interval of the Probability Density Function over each X,Y,Z coordinate values. & GAPI-PDF-X, GAPI-PDF-Y, GAPI-PDF-Z.\\
8 & GAPI-CDF & Average of peak-to-peak interval of the Cumulative Distribution Function over each X,Y,Z coordinate values. & GAPI-CDF-X, GAPI-CDF-Y, GAPI-CDF-Z.\\
9 & FFT-ADPI-CDF & Average of intervals between distinct peaks of the Fast Fourier Transformation of the Cumulative Distribution Function for X,Y,Z coordinates. & FFT-ADPI-CDF-X, FFT-ADPI-CDF-Y, FFT-ADPI-CDF-Z. \\	
\hline
\end{tabular}
\end{center}
\end{table*}

\section{Experiments}
\label{section:Experiments}

We performed experiments by learning three different Machine Learning (ML) classifiers (algorithms) using the WEKA ML platfom with default settings \cite{hall2009}. The classifiers consist of, Sequential Minimum Optimization (SMO), which is a variant of Support Vector Machines (SVM), Na\"{i}ve Bayes (NB), and Neural Networks with Multilayer Perceptron (NN-MP).

For learning the classifiers, we used a meta classification approach that combines attribute selection techniques to reduce the features space while discarding the less contributing features. As such, we used the InfoGain attribute selection algorithm, together with the Ranker algorithm \cite{makrehchi2012}.

Our dataset\footnote{See dataset on https://github.com/soori1/NearnessRecognition} consists of 1000 instances comprising of 500 instances for steady-space (labeled \emph{control}), 500 instances for non-steady space (labeled \emph{near}). 

\subsection{Results and Discussion}
\label{section:ResultsDiscussion}

In the first stage, we evaluated the prediction performance of the three ML classifiers on the \emph{control} and \emph{near} categories with random 60\% training and 40\% testing set, see Table \ref{table:performance-all-sensors}.

\begin{table*}[ht]
\begin{center}
\caption{\label{table:performance-all-sensors}60\%-40\% classifiers' performance on Steady (Control) and Non-Steady (Near) Spaces using all 9 sensors.}
\begin{tabular}{p{1.5cm} p{1.8cm} p{1.5cm} p{1cm} p{1.8cm}} 
\hline
\textbf{Category}&\textbf{Classifier}&\textbf{Precision}&\textbf{Recall}&\textbf{F-Measure}\\ 
\hline
Near & SVM-SMO & 0.8 & 0.788 & \textbf{0.794}\\ 
Control & SVM-SMO & 0.795  &   0.807   &  \textbf{0.801}\\ 
Near & NB & 0.557  &   0.884   &  0.684\\ 
Control & NB & 0.733  &   0.312  &   0.438\\ 
Near & NN-MP & 0.612   &  0.702   &  0.654\\ 
Control & NN-MP & 0.659  &   0.564   &  0.608\\
\hline
\end{tabular}
\end{center}
\end{table*}

In the second stage of our experiment, we evaluated the contributions of each sensor to predicting the \emph{near} category with the SVM-SMO classifier using a 60\%-40\% split on the 1000 instances. This is to know which sensor(s) are more effective for prediction. As such, we removed each sensor's readings one at a time and then classify with every other sensors, see  Table \ref{table:performance-each-sensor}. 

\begin{table*}[ht]
\begin{center}
\caption{\label{table:performance-each-sensor}SVM-SMO performance on the contributions of each sensor to predicting the \emph{near} category.}
\begin{tabular}{p{3.5cm} p{1.5cm} p{1cm} p{1.8cm} p{2cm}} 
\hline
\textbf{Model}& \textbf{Precision}&\textbf{Recall}&\textbf{F-Measure}&\textbf{Baseline Diff.}\\ 
\hline
Without-light &  0.801  &   0.773  &   0.787 &   -0.005 $\downarrow$\\
Without-gravity & 0.75  &    0.844  &   0.794&   0\\
Without-accelerometer  & 0.786  &   0.77   &   0.778 &   -0.016 $\downarrow$\\
Without-rotation-vector  & 0.724  &   0.844   &  0.779 &   -0.015 $\downarrow$\\
Without-temperature & 0.773  &   0.757   &  0.765 &   -0.029 $\downarrow$\\
\textbf{SVM-SMO Baseline-all sensors} & 0.8 & 0.788 & \textbf{0.794} &   N/A\\
Without-humidity & 0.795  &   0.852   &  0.823 &  \textbf{0.029} $\uparrow$\\
Without-gyroscope  & 0.844  &   0.824   &  0.834 &   \textbf{0.04} $\uparrow$\\
Without-magnetometer  & 0.826  &   0.874   &  0.849 &  \textbf{0.055} $\uparrow$\\
Without-pressure  & 0.852   &  0.877   &  0.864 &  \textbf{0.07} $\uparrow$\\
\hline
\end{tabular}
\end{center}
\end{table*}

Using the SVM-SMO 79.4\% F-measure as baseline (see Table \ref{table:performance-all-sensors}), we identified which sensor did not improve the performance of the classifier. If a sensor is removed, and the performance of the model increases more than 79.4\% F-measure, then that sensor is considered as not improving the performance of the model.

Further, we evaluated by using SVM-SMO with 60\%-40\% split to compare the performance of the model with all the 9 sensors to the model with just the 5 sensors that gave better performance (i.e. light, gravity, accelerometer, rotation-vector, and temperature). We then compute the \emph{sensitivity} and \emph{1-specificity} of the two models for comparison. Table \ref{table:sens-spec} shows the comparison for the two models. The 5-sensor model clearly showed better sensitivity of 88.3\% for performing the nearness recognition task \cite{haron:2012,mackowiak:2011}.

\begin{table*}[ht]
\begin{center}
\caption{\label{table:sens-spec}Sensitivity and 1-Specificity comparison between the 5-sensor and all-sensors models.}
\begin{tabular}{p{2cm} p{2cm} p{2cm} p{2cm} p{2.2cm}} 
\hline
\textbf{Model}& \textbf{Accuracy\%}&\textbf{Sensitivity\%}&\textbf{Specificity\%}&\textbf{1-Specificity\%}\\ 
\hline
5-Sensor &  \textbf{88.75}  &   \textbf{88.3}  &   89.2	 &   \textbf{10.8}\\
All sensors & 79.75  &    78.8  &   80.7 &   19.3\\

\hline
\end{tabular}
\end{center}
\end{table*}

\begin{figure}[ht]
\begin{center}
\includegraphics[width=65mm,height=35mm]{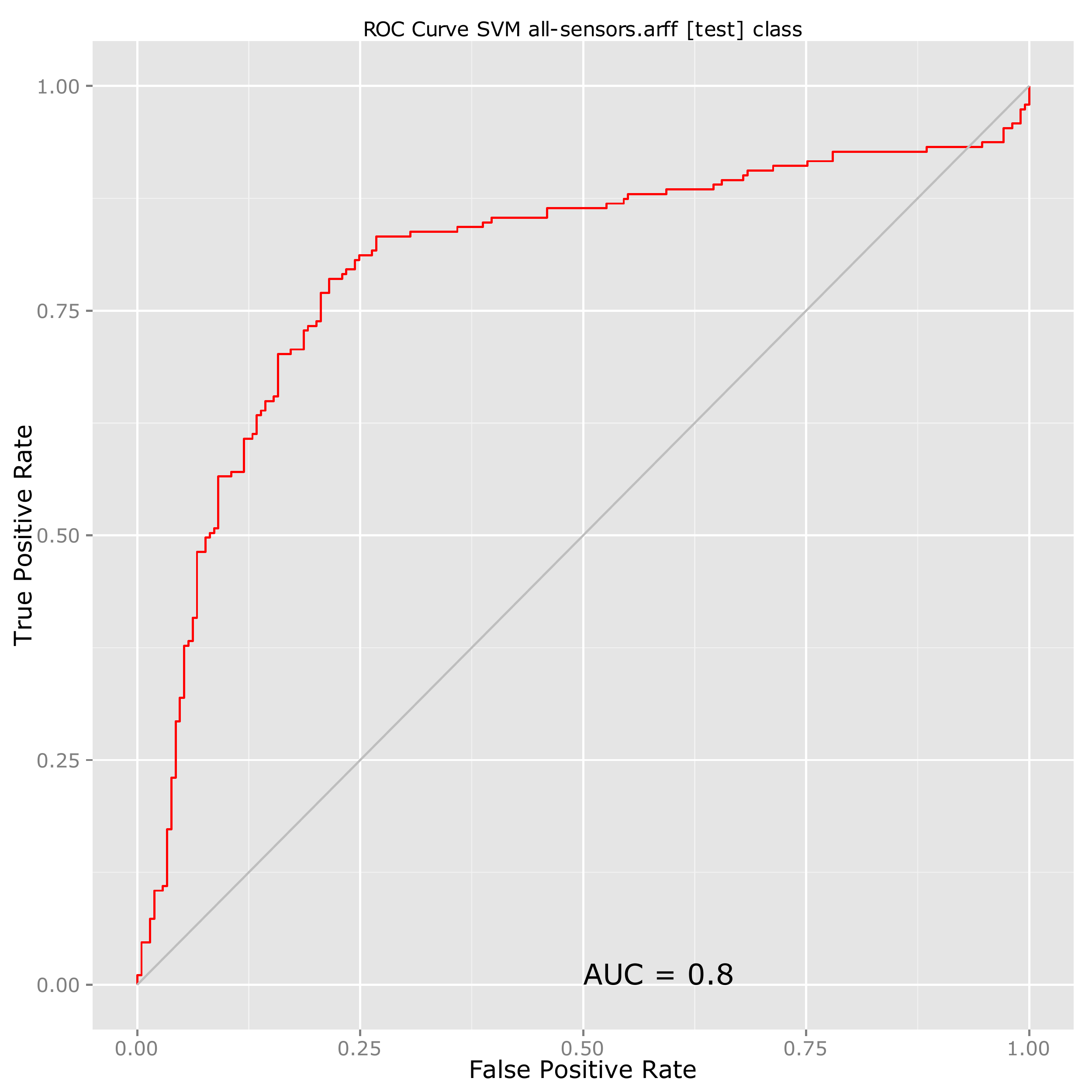}
\caption{\label{fig:roc-all-sensors} ROC Curve (AUC = 0.8) for SVM-SMO 9 sensors.}
\end{center}
\end{figure}

\begin{figure}[ht]
\begin{center}
\includegraphics[width=65mm,height=35mm]{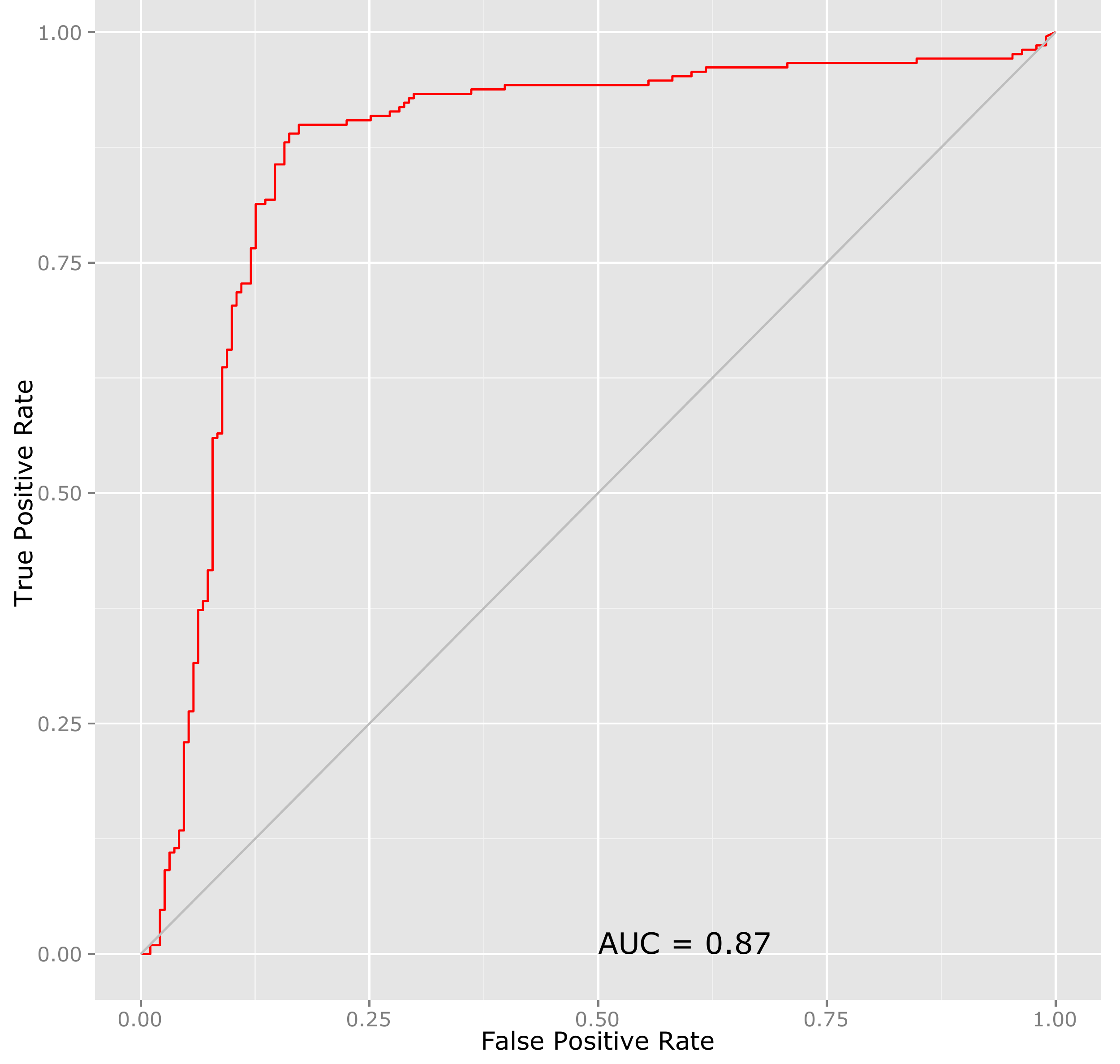}
\caption{\label{fig:roc-good-sensors} ROC Curve (AUC = 0.87) for SVM-SMO 5 best sensors.}
\end{center}
\end{figure}

Furthermore, we compared between the Receiver Operator Characteristic (ROC) curves of the two models \cite{majnik2013}. Figure \ref{fig:roc-all-sensors} shows the ROC curve for a model with all 9 sensors and Figure \ref{fig:roc-good-sensors} shows the ROC curve for a model with the 5 sensors. The 5-sensor model shows better Area Under Curve (AUC) of 0.87 compared to 0.80 for 9 sensors.

\section{Conclusion and Future Work}
\label{section:ConclusionFutureWork}

We performed human-mobile nearness recognition for non-invasive health diagnostic purposes by analyzing and learning data from 9 different sensors on a mobile device. Several data transformation were done to learn the unique characteristics of both \emph{steady} and \emph{non-steady} three-dimensional spaces for predicting human nearness to a mobile device. The limitation of this work lies in the fact that test data consists of sensor readings that are limited to a controlled environment. It could be better to consider several uncontrolled environments including a public space which might contain several ambient noise signals at different weather and day times. Also, it could be challenging to execute the complex state-of-the-art ML algorithms on the devices. As such, collected ambient sensors’ data could be sent to a remote server in real-time, where the trained model can be used for prediction and then send the output back to the device. In the future, we plan to detect human breathing rate using the same technique and include a diagnostic web interface for medical practitioners.

\section*{Acknowledgement}
This project was funded by the VITAL and the Tropical Medicine and Biology grants of Monash University Malaysia.


\bibliographystyle{IEEETran}
\small
\bibliography{references}

\end{document}